\documentclass[a4paper]{article}

\usepackage{itgspeech2018}    
\usepackage{times}            
\usepackage[english]{babel}   
\usepackage[ansinew]{inputenc}
\usepackage[T1]{fontenc}      
\usepackage[sort&compress,numbers]{natbib}	
\usepackage{amsmath,amssymb}
\usepackage{graphicx}
\usepackage[colorlinks=false,pdfborder={0 0 0}]{hyperref}
\usepackage{units}
\usepackage{multirow}
\usepackage{tikz}
\usepackage{pgfplots}
\usepackage{subcaption}

\title{Unsupervised Domain Adaptation by Adversarial Learning for Robust Speech Recognition}

\author{Pavel Denisov, Ngoc Thang Vu, Marc Ferras Font}

\address{Institute for Natural Language Processing, University of Stuttgart, Germany\\
  Email: \texttt{\{pavel.denisov,thang.vu,marc.ferras\}@ims.uni-stuttgart.de}}

\begin{document}

\maketitle

\begin{abstract}
In this paper, we investigate the use of adversarial learning for unsupervised adaptation to
unseen recording conditions, more specifically, single microphone far-field speech. 
We adapt neural networks based acoustic models trained with close-talk clean speech to the new recording
conditions using untranscribed adaptation data. 
Our experimental results on Italian SPEECON data set show that our proposed method achieves 19.8\% relative word error rate (WER) reduction
compared to the unadapted models.
Furthermore, this adaptation method is beneficial even when
performed on data from another language (i.e. French) giving 12.6\% relative WER reduction.
\end{abstract}


\section{Introduction}

Recently with the success of deep learning methods, automatic speech recognition (ASR) has achieved human performance in conversational speech recognition \cite{xiong2016achieving, xiong2017toward}.
However, far-field speech, especially when recorded with single microphone, remains one of
the major obstacles to achieving complete human parity,
mainly because of challenging environments with a lot of noises and reverberations
\cite{peddinti2016far, ko2017study, zhang2017attention, yi2018distilling, spille2018comparing}.
Another challenge is that it is almost impossible to collect data covering all recording environments to train and to test on
due to variations of reverberations/noises and distances to microphones.
While there were some advancements in this direction for a few
widespread languages, a large number of low-resource languages
will inevitably remain uncovered by such kind of resources.
These facts motivate our interest for methods
to improve robustness of acoustic model by utilizing
available data, especially data from resource rich languages \cite{vu2014automatic}.

It is well known that a mismatch between training and testing
conditions is likely to degrade accuracy of acoustic models.
In case of Deep Neural Network (DNN) acoustic modeling, this issue
can be addressed by the wide range of transfer learning methods
developed by the deep learning community.
Two well researched transfer learning methods in ASR are
weights transfer and multi-task learning.
These two closely related methods were shown to be efficient in case
of language mismatch \cite{vu2013multilingual, heigold2013multilingual, kunze2017transfer},
recording conditions mismatch \cite{ghahremani2017investigation}
and combination of them \cite{zhuang2017improving}.

Common ways to improve accuracy of acoustic models on
speech with different noises and reverberation strengths and forms,
as well as used recording equipment, are
speech enhancement techniques \cite{zhao2017two},
feature engineering \cite{kim2012power, ganapathy2017multivariate, ichikawa2017harmonic},
simultaneous training on recordings from different environments,
particularly from simulated ones \cite{kim2017generation},
and such standard deep learning techniques as dropout \cite{kovacs2017increasing}.
Domain adaptation is another form of transfer learning,
which is often used to perform speaker adaptation in ASR,
but can also be applied to solve the problem of recording conditions mismatch.
Adversarial multi-task learning was proposed for domain adaptation
in image classification \cite{ganin2014unsupervised} and later shown to be efficient
on other tasks \cite{ganin2016domain}.

The goal of adversarial multi-task learning is to combine
feature extraction and domain adaptation problems to
the single training process in a such way that learned features
are discriminant to the main task and invariant to the domain.
This is achieved by joint optimization for two objectives
with well known multi-task training method, which involves
sharing of some lower DNN layers between tasks
and employing the output of the last shared layer
as implicitly learned features for the
two smaller task specific sub-networks.
In adversarial multi-task learning,
one of the tasks is the main classification task
and another one is the domain discrimination task.
Gradient of the loss function is propagated to the shared layers
as usual from the main classification task;
however, gradient of the domain classification loss function
is inversed before being propagated to the shared layers,
what promotes minimization of domain classification
accuracy and domain invariance of the output of the last
shared layer, thus making the input of
the main classifier discriminant to the main task
and domain invariant at the same time.

Recent works investigated application of this method to improve the robustness of ASR
to variations in noise types and levels \cite{shinohara2016adversarial, serdyuk2016invariant, sun2017unsupervised},
to accented speech \cite{sundomain} and to speaker variation \cite{meng2018speaker}.
We evaluate applicability of adversarial learning for
unsupervised adaptation of an acoustic model
trained on clean close-talk speech to speech recorded with a single microphone and in a far-field scenario.
We explore both cases, within the language (Italian) and language mismatch (Italian as target language and adaptation data is French) 
and compare its efficiency with other supervised training methods.
To the best of our knowledge, this study is the first evaluation of unsupervised domain adaptation for ASR in a crosslingual setup.


\section{Method}
\label{sec:method}
\begin{figure}[ht]
  \centerline{\includegraphics[width=\columnwidth,trim={0 6cm 0 0},clip]{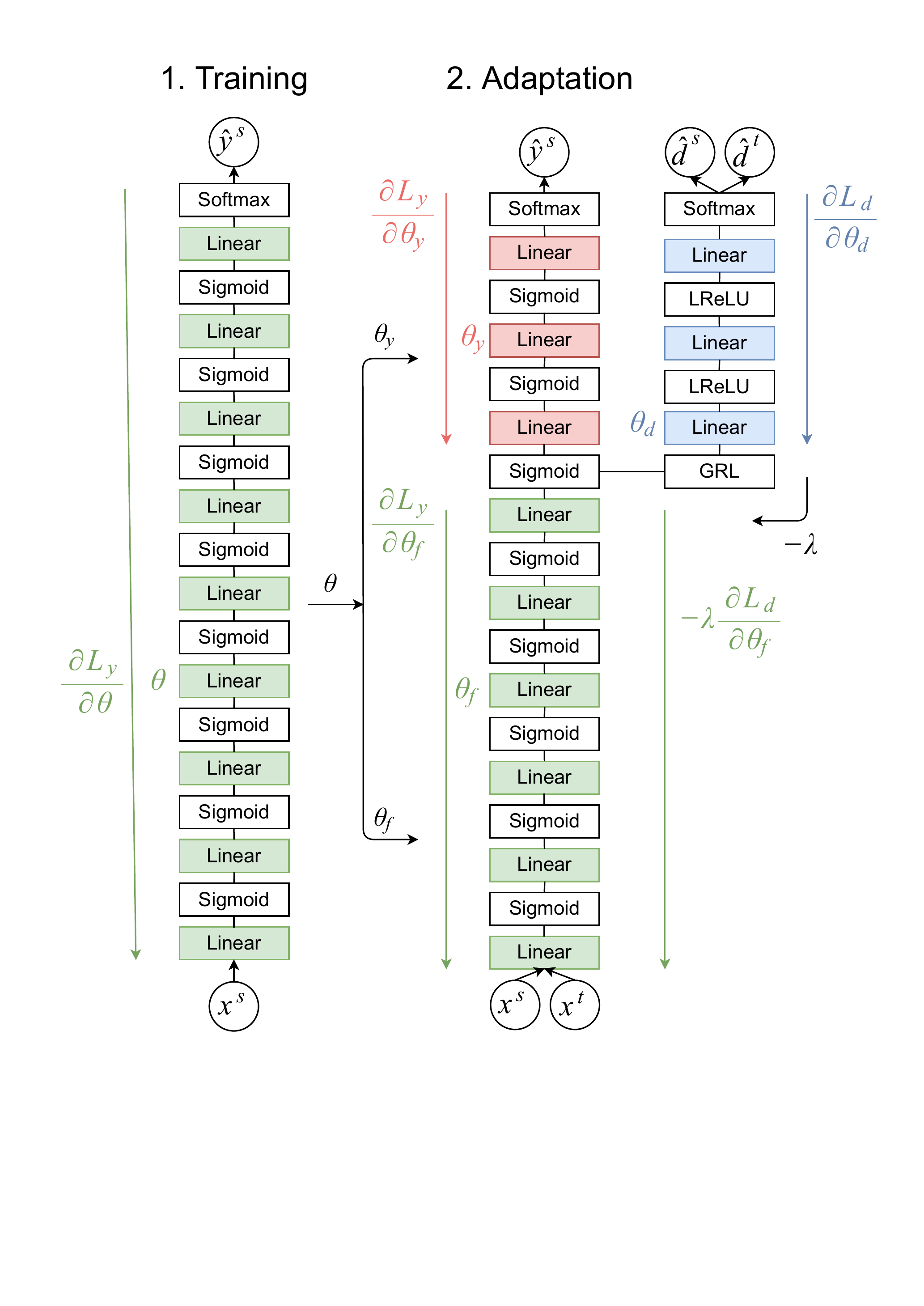}}
  \caption{The architecture of DNN during the training and adaptation stages.}
  \label{fig:mlp_amtl_unsupervised}
\end{figure}

Inspired by \cite{meng2017unsupervised}, we first perform regular
training of DNN acoustic model on labeled source domain data (clean speech) and then
adapt learned weights using mixture of same labeled source domain data and
unlabeled target domain data (noisy reverberated speech). An overview of the method
is shown in Fig.~\ref{fig:mlp_amtl_unsupervised}.

At the training stage, we only use data samples from the source domain
$X^s = \{ x_1^s, \dots, x_{N^s}^s \}$ and corresponding
senone labels $Y^s = \{ y_1^s, \dots, y_{N^s}^s \}$. Based on DNN
parameters $\theta$, we calculate predicted
senone labels $\hat{Y}^s = \{ \hat{y}_1^s, \dots, \hat{y}_{N^s}^s \}$
 and the value of cross-entropy loss function
\begin{equation}
L_y(\theta) = - \frac{1}{N^s} \sum_{i = 1}^{N^s} \log P(\hat{y}_i^s = y_i^s|x_i^s;\theta),
\end{equation}
The parameters are then updated via back-propagation
for minimization of the loss function:
\begin{equation}
\theta \gets \theta - \epsilon \frac{\partial L_y}{\partial \theta},
\end{equation}
where $\epsilon$ is the learning rate.

At the adaptation stage, we use the same data samples from the source domain
$X^s = \{ x_1^s, \dots, x_{N^s}^s \}$ and corresponding
senone labels $Y^s = \{ y_1^s, \dots, y_{N^s}^s \}$.
We also add data samples from the target domain $X^t = \{ x_1^t, \dots, x_{N^t}^t \}$,
for which we do not have senone labels.
In addition to that, we introduce secondary task of domain classification.
The set of parameters $\theta$, which were learned at the training stage,
is decomposed into two sets: the parameters of the first $f$ hidden layers $\theta_f$,
which are shared between the senone and domain classification tasks and act as a feature extractor,
and the rest of the parameters $\theta_y$, which are used by senone classification part of DNN.
New set of parameters $\theta_d$ is added for the domain classification task.
Loss functions for the senone and domain classification tasks are defined as follows:
\begin{equation}
L_y(\theta_f, \theta_y) = - \frac{1}{N^s} \sum_{i = 1}^{N^s} \log P(\hat{y}_i^s = y_i^s|x_i^s;\theta_f, \theta_y)
\end{equation}
\begin{equation}
\begin{split}
L_d(\theta_f, \theta_d) = - \frac{1}{N^s} \sum_{i = 1}^{N^s} \log P(\hat{d}_i^s = 1|x_i^s;\theta_f, \theta_d) \\
- \frac{1}{N^t} \sum_{i = 1}^{N^t} \log P(\hat{d}_i^t = 2|x_i^t;\theta_f, \theta_d)
\end{split}
\end{equation}
Note that we do not take into account senone label predictions for the target domain data samples
$X^t$, because we do not know true senone labels for them.
Task specific parameters are updated to minimize corresponding loss functions:
\begin{equation}
\theta_y \gets \theta_y - \epsilon \frac{\partial L_y}{\partial \theta_y}
\end{equation}
\begin{equation}
\theta_d \gets \theta_d - \epsilon \frac{\partial L_d}{\partial \theta_d}
\end{equation}
The update of shared parameters is performed so that it minimizes the senone classification
loss function and maximizes the domain classification loss function:
\begin{equation}
\theta_f \gets \theta_f - \epsilon \left( \frac{\partial L_y}{\partial \theta_f}
- \lambda \frac{\partial L_d}{\partial \theta_f} \right)
\end{equation}
Maximization of the domain classification loss function aims making the output
of the last shared hidden layer as less informative for the domain classifier
as possible, and thus similar for data samples from different domains.
The negative coefficient $-\lambda$ is responsible for that and for the balance between
the importance of this task and the primary task of senone classification.
$\lambda$ is initially set to $0$ and is increased gradually
in the training process according to the following function:
\begin{equation}
\lambda_e = \min(\frac{e}{10}, 1) \lambda,
\end{equation}
where $\lambda_e$ is the value of gradient reversal coefficient used during epoch $e$.
That is done in order to allow the senone classification part of DNN
to adjust its parameters to the output of the feature layer,
which would be changed too fast by the
domain classification part of DNN otherwise.

\section{Experimental Setup}
\label{sec:experiments}
\subsection{Datasets}
SPEECON is a family of speech corpora purposed for the development of
speech recognition in consumer devices. The corpora were recorded for many
languages according to the common specifications what allows us to evaluate the propose method in case of language mismatch while other
conditions are not altered in a significant way.
We use Italian data set for all the experiments and French as adaptation data in the cross-lingual experiment.
Each corpus contains recordings of read and spontaneous speech
by 550 adult speakers.
The recordings are made in four acoustic environments: office, entertainment, public place and car.
Each recording is made with 4 microphones located on different distances
from the speaker that are represented by 4 channels in SPEECON corpora:
\begin{itemize}
\item Channel 1 corresponds to a close distance headset microphone placed right in front of the speaker's mouth;
\item Channel 2 corresponds to a lavalier microphone placed below the chin of the speaker;
\item Channel 3 corresponds to a middle distance microphone placed in 0.5--1.0 meters from the speaker;
\item Channel 4 corresponds to a far distance omni-directional microphone in office and entertainment
environments or middle distance otherwise.
\end{itemize}
Transcriptions are converted to lower case and cleaned up from punctuation marks.
Summary of the used corpora is given in Tab.~\ref{tab:speecon}.

In addition to that, 197 millions words of Italian Deduplicated CommonCrawl Text are used to build Italian language model. 
Italian dictionary ILE with pronunciations for 588k words is used as a lexicon.

\begin{table}[h]
  \centering
  \begin{tabular*}{0.8\columnwidth}{l|l|l}
    \noalign{\hrule height 1pt}
     & Italian & French \\
    \hline
    Number of utterances & 174,940 & 182,679 \\
    Number of speakers & 451  & 495 \\
    Duration (hours) & 157 & 167 \\
    \noalign{\hrule height 1pt}
  \end{tabular*}
  \caption{ Summary of SPEECON corpora. }
  \label{tab:speecon}
\end{table}

\subsection{Baseline}
Our DNN-Hidden Markov Model (HMM) acoustic model is a multilayer perceptron
consisting of 8 hidden fully connected layers with 1024 units each and
output layer with 9315 units corresponding to senones (HMM states).
Sigmoid activation function is used for the hidden layers
and softmax activation function is used for the output layer.
We use Adam optimizer \cite{kingma2014adam}
and new-bob learning rate scheduler \cite{morgan1995continuous}
with initial learning rate of 0.0001 for training.
The input of DNN is 23-band log Mel filterbank features with delta and delta-deltas
and splicing with 5 context frames both left and right, giving 759 dimensions in total.
Training process iterates over data samples in randomized order with mini-batch size of 256 samples.
NNabla \cite{nnabla} deep learning toolkit is used to implement DNN.
Kaldi speech recognition toolkit \cite{povey2011kaldi} is used to build Gaussian Mixture Model-HMM acoustic model,
to produce forced senone level alignments of training data required for DNN-HMM training
and to perform decoding with DNN-HMM required for WER evaluation.
For decoding we also trained two 3-gram language models on the transcripts from the training data
and on the CommonCrawl subset and interpolated them with SRILM toolkit \cite{stolcke2002srilm}.
The perplexity of the language model on our testing data set is 209.47.

Results of the baseline model trained on different data sets with different labels
and tested on 15 hours of Italian SPEECON channel 4
are shown in Tab.~\ref{tab:baseline}.
It is apparent from this table that decoding of noisy reverberated speech is a challenging task.
While WER of the model trained on Channel 1 is incredibly high at 85.2\% due to significant
distortions introduced to speech by environmental noises
and reverberations in the testing noisy speech data,
the model trained on Channel 4 achieves significantly
lower WER by learning to normalize these distortions from the training data,
and the model trained on Channels 1--4 results even better WER
because of the generalized representations of clean and noisy
data samples presenting in the training data.
Our analysis of problematic utterances suggests that
as the majority of the mistakes are made in
,,Spontaneous speech'', ,,Numbers, times, dates'' and ,,Named entities'' categories,
where the language model could not be helpful.

An alternative to the proposed method would be
to train a model on the target domain data
and the labels produced by a first pass of
unadapted model. As it follows from Tab.~\ref{tab:baseline},
this method does not seem to be practical in our setup,
most likely because of extremely bad accuracy
of the unadapted model. Moreover, the proposed method
has an advantage of applicability in a crosslingual setup.

\begin{table}[h]
  \centering
  \begin{tabular*}{0.7\columnwidth}{l|l|r}
    \noalign{\hrule height 1pt}
    System & Training data & WER (\%)\\
    \hline
    Baseline & Channel 1 & 85.2 \\
    First pass & Channel 4 & 86.3 \\
    Oracle & Channel 4 & 51.8 \\
    Oracle & Channels 1--4 & 46.0 \\
    \noalign{\hrule height 1pt}
  \end{tabular*}
  \caption{ Results of the baseline model. }
  \label{tab:baseline}
\end{table}

\subsection{Setup Description}
Each of the experiments starts with DNN weights trained on 125 hours of clean close-talk Italian speech training data.
Adaptation stage is performed on a combination of clean speech training data with senone labels
and noisy speech adaptation data without senone labels (technically they all are set to $0$).
The domain classification sub-network is added at adaptation stage
and consists of 2 hidden fully connected layers with 512 units each and
the output layer with 2 outputs
corresponding to the source and target domain classes in the adversarial task.
Leaky ReLU activation function \cite{maas2013rectifier} is used for the hidden layers and
softmax activation function is used for the output layer.
Input of domain classification sub-network is output
of the $f$-th hidden layer of the main network (feature layer) passed through
a Gradient Reversal Layer (GRL).
GRL passes its input intact to its output
during the forward pass and
returns the inversed and scaled by $\lambda$ gradient value
from its output to its input during the backward pass.
Adaptation procedure could be then interpreted as a regularizer of the DNN training.
After it is finished, the domain classification sub-network is removed
and decoding is performed as usual with the remaining DNN.

Three experiments are conducted to investigate the
effectiveness of the proposed method.
In the first experiment, we investigate the interaction between GRL coefficient $\lambda$ and feature layer index $f$ using 125 hours of channel 4 (middle/far distance microphone) as adaptation data.
The best GRL coefficient $\lambda$ and feature layer index $f$ are then used for further experiments.
The second experiment explores the impact of the adaptation data size on the final performance. 
In the third experiment, we perform a cross-lingual study when using the same amount of adaptation data but from French in order to examine importance of the language of adaptation data.

\section{Results}
\label{sec:results}

\subsection{Training Metrics}

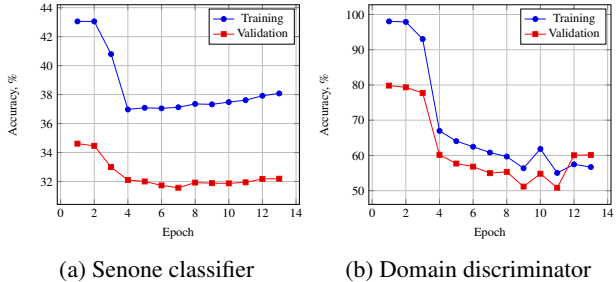
\begin{figure}[h]
  \centering
  \begin{subfigure}[b]{.5\columnwidth}
  \centering
  \resizebox{\textwidth}{!}{
  \begin{tikzpicture}
\begin{axis}[
	grid=major,
	ylabel={Accuracy, \%},
	xlabel={Epoch},
]
	
\addplot coordinates {
(1,43.0526)
(2,43.0525)
(3,40.7944)
(4,36.9719)
(5,37.0871)
(6,37.0474)
(7,37.1242)
(8,37.3525)
(9,37.3246)
(10,37.4797)
(11,37.6131)
(12,37.9185)
(13,38.0779)
};
\addlegendentry{Training}

\addplot coordinates {
(1,34.6112)
(2,34.461)
(3,32.9953)
(4,32.0962)
(5,32.001)
(6,31.7289)
(7,31.5663)
(8,31.9229)
(9,31.8834)
(10,31.8698)
(11,31.9409)
(12,32.176)
(13,32.1885)
};
\addlegendentry{Validation}

\end{axis}
\end{tikzpicture}
  }
  \caption{Senone classifier}
  \end{subfigure}%
  \begin{subfigure}[b]{.5\columnwidth}
  \centering
  \resizebox{\textwidth}{!}{
  \begin{tikzpicture}
\begin{axis}[
	grid=major,
	ylabel={Accuracy, \%},
	xlabel={Epoch},
]
	
\addplot coordinates {
	(1,98.03804864027195)
	(2,97.88464182163567)
	(3,93.0213176114777)
	(4,66.96840319436113)
	(5,64.05242389022196)
	(6,62.46414779544091)
	(7,60.81322797940412)
	(8,59.6810325434913)
	(9,56.35536955108978)
	(10,61.844193661267745)
	(11,55.03157181063787)
	(12,57.4593675014997)
	(13,56.70444036192762)
};
\addlegendentry{Training}

\addplot coordinates {
	(1,79.81894241841395)
	(2,79.32727402501676)
	(3,77.7175443202956)
	(4,60.167781857782785)
	(5,57.691403135953536)
	(6,56.81491768636036)
	(7,54.97412573392638)
	(8,55.34022885374619)
	(9,51.15229110778595)
	(10,54.80496380983715)
	(11,50.82358126020221)
	(12,60.043328743657234)
	(13,60.14016570286105)
};
\addlegendentry{Validation}

\end{axis}
\end{tikzpicture}
  }
  \caption{Domain discriminator}
  \end{subfigure}
  \caption{Accuracy during the adaptation stage.}
  \label{fig:adaptation_accuracy}
\end{figure}

Fig.~\ref{fig:adaptation_accuracy}
shows  the accuracy values for senone classification and domain discrimination
during the adaptation stage.
Training data set consists of equal proportions of Channel 1 and Channel 4 recordings
and validation data set consists of equal proportions of Channels 1--4 recordings of Italian SPEECON corpus.
Accuracy is defined as the number of correctly classified samples divided by the total number of samples.
What stands out here is the markedly high accuracy of the domain classifier
during the initial epochs, which suggests that the feature layer initially outputs
quite distinct values for the clean and noisy speech. As the GRL coefficient is increased
and the shared DNN parameters are adjusted towards more domain invariant representation,
the accuracy of the domain classifier expectedly decreases and stabilizes
slightly over the chance level around 55\%. At the same time, the senone classification
accuracy first drops quite sharply in response to changes in how the feature layer
represents the data and later recovers slowly due to the adaptation
of the task specific layers to the new domain invariant output of the feature layer
made possible by the utilization of the labeled clean speech data samples.
Another interesting observation can be made by comparing the metrics
of the domain classifier for the training and validation data sets.
The performance of the domain classifier for the training and validation data sets aligns to similar level after a few epochs
of adaptation, which indicates that the representation learned by the shared DNN parameters
does not just normalize seen data samples, but actually extracts
only the information not related with recording conditions.

\subsection{Effect of $\lambda$ and $f$}
\begin{table}[h]
  \centering
  \begin{tabular*}{0.455\columnwidth}{l|c|c|c}
    \noalign{\hrule height 1pt}
    \multirow{2}{*}{$f$} & \multicolumn{3}{c}{$\lambda$} \\\cline{2-4}
      & 1.0 & 2.0 & 4.0 \\
    \hline
    1 & 84.5 & 78.7 & 76.8  \\
    2 & 70.2 & \textbf{68.3} & 69.0 \\
    3 & 69.8 & 69.2 & 74.5 \\
    4 & 71.9 & 71.5 & 74.9 \\
    5 & 74.1 & 74.7 & 75.6 \\
    \noalign{\hrule height 1pt}
  \end{tabular*}
  \caption{ Results (WER in \%) of adaptation on 125 hours of Channel 4. }
  \label{tab:lambda_feature}
\end{table}

First we use the fixed target domain data subset, namely 125 hours of Channel 4 recordings,
to evaluate the effect of various combinations of the gradient reversal coefficient $\lambda$ and
the feature layer index $f$. WER and relative error rate reduction (RERR)
are listed in Tab.~\ref{tab:lambda_feature}.
The best combination is gradient reversal coefficient $\lambda = 2.0$ and
feature layer index $f = 2$ and it results WER of 68.3\%, which is
within almost twice smaller gap with the best result of 46.0\%,
obtained by supervised training on Channels 1--4, compared
to 85.2\% resulted by unadapted model trained on Channel 1.
In addition to that, we repeat the same experiment with Channels 2--4
as adaptation data, as the best baseline system also utilizes this data.
We obtain WER of 66.6\% and conclude that the gain in WER is too small
in comparison to amount of additional adaptation data.

\subsection{Effect of Adaptation Data}
\begin{table}[h]
  \footnotesize
  \centering
  \begin{tabular*}{0.5\columnwidth}{l|l|l}
    \noalign{\hrule height 1pt}
	Hours & WER (\%) & RERR (\%) \\
    \hline
    - & 85.2 & - \\
    50 & 69.7 & 18.2  \\
    40 & 70.5 & 17.2 \\
    30 & 71.8 & 15.7 \\
    20 & 75.4 & 11.5 \\
    10 & 85.4 & -2.3 \\
    5 & 84.0 &  1.4 \\
    \noalign{\hrule height 1pt}
  \end{tabular*}
  \caption{ Results of adaptation on Italian data with $f = 2$, $\lambda = 2.0$. }
  \label{tab:adaptation_data_italian}
\end{table}

Next we use the best combination of the gradient reversal coefficient $\lambda$ and
the feature layer index $f$ from the previous experiment to evaluate contribution
of various amounts of adaptation data to accuracy of adapted model.
Results are listed in Tab.~\ref{tab:adaptation_data_italian}.
What emerges from the results reported here is that no significant
drop in WER is observed if amount of adaptation data is decreased
to 30 hours or one third of originally evaluated adaptation data set.
On the one hand this finding suggests that one does not need to acquire
large amount of the target domain data in order
to get a moderate improvement of ASR system trained on clean speech data.
On the other hand, it is possible that this effect of 30 hours of adaptation data
is due to a good chance of having comparable number of distinct recording conditions
in the adaptation and testing data
and may not be generalizable to a
larger testing data set with more diverse set of recording conditions.

\subsection{Crosslingual Adaptation}
\begin{table}[h]
  \footnotesize
  \centering
  \begin{tabular*}{0.5\columnwidth}{l|l|l}
    \noalign{\hrule height 1pt}
	Hours & WER (\%) & RERR (\%) \\
    \hline
    - & 85.2 & - \\
    50 & 74.5 & 12.6 \\
    40 & 75.6 & 11.3 \\
    30 & 76.4 & 10.3 \\
    20 & 80.2 & 5.9 \\
    10 & 83.7 & 1.8 \\
    5 & 84.4 & 0.9 \\
    \noalign{\hrule height 1pt}
  \end{tabular*}
  \caption{ Results of adaptation on French data with $f = 2$, $\lambda = 2.0$. }
  \label{tab:adaptation_data_french}
\end{table}

We also run experiments on the same amounts of French data to see if it is important to use adaptation data
for the same language as the language of interest.
Results are listed in Tab.~\ref{tab:adaptation_data_french}.
Interestingly, the method improves WER
even when used with the adaptation data for French while
the language of interest is Italian. We also observe the same trend
regarding amount of French adaptation data as with Italian
adaptation data, namely insignificant contribution
of additional adaptation data, besides 30 hours, to WER.
Hence, it could conceivably be hypothesized that
the method makes DNN more robust to a number of different recording
conditions in general and not only to the recording conditions represented
in the adaptation data.

\section{Conclusions}
\label{sec:conclusions}
The present study was designed to gain a better understanding of
ability of unsupervised domain adaptation by adversarial Learning
to improve robustness of ASR. We perform
adaptation experiments on close-talk and far/middle distance recordings using the Italian and French SPEECON corpora.
Our experimental results show that the proposed method improved significantly the WER in case
of recording conditions mismatch without any transcriptions.
Up to 19.8\% relative WER improvement could be observed.
Additionally, results on cross-lingual experiments also indicate that the usage of adaptation data
from the same language is desirable, but not mandatory. Adaptation on French data resulted relative WER improvement up to 12.6\%.

The present investigation has not considered more distant
pairs of languages having smaller overlap in phonetic inventory,
which is one of possible directions for the future research.
Further work needs to be done to establish whether our conclusions
would hold for more advanced DNN architectures, such as TDNN \cite{waibel1990phoneme, peddinti2015time}, LSTM \cite{graves2013speech}
and CNN \cite{abdel2012applying}, and training methods, such as Lattice-free MMI \cite{povey2016purely}.

\small
\bibliographystyle{ieeetr}
\bibliography{paper}


\end{document}